\documentclass{svjour3}
\smartqed  
\usepackage{amsmath}
\usepackage{amsopn}
\usepackage{graphicx}
\usepackage[breaklinks=true,colorlinks=true,linkcolor=blue,urlcolor=blue,citecolor=blue]{hyperref}
\newcommand{\bea}{\begin{eqnarray}}
\newcommand{\eea}{\end{eqnarray}}
\newcommand{\be}{\begin{equation}}
\newcommand{\ee}{\end{equation}}
\newcommand{\nn}{\nonumber}

\newcommand{\la}{\langle}
\newcommand{\ra}{\rangle}
\newcommand{\ah}{\hat{a}}
\newcommand{\ahd}{\hat{a}^\dag}
\newcommand{\bj}{\hat{b}_j}
\newcommand{\bjd}{\hat{b}^\dag_j}
\newcommand{\bk}{\hat{b}_k}
\newcommand{\bkd}{\hat{b}^\dag_k}
\newcommand{\oj}{\omega_j}
\newcommand{\ok}{\omega_k}
\newcommand{\Uh}{\hat{U}}
\newcommand{\Uhd}{{\hat{U}}^\dag}
\newcommand{\roh}{\hat{\rho}}
\journalname{European Physical Journal Plus}
\begin{document}
\title{Open quantum systems in Heisenberg picture}
\author{Fardin Kheirandish}

\institute{ Fardin Kheirandish\at
              Department of Physics, Faculty of Science, University of Kurdistan, P.O.Box 66177-15175, Sanandaj, Iran \\
              Tel.: +98-87-33668332\\
              \email{f.kheirandish@uok.ac.ir}}                   

\institute{Department of Physics, Faculty of Science, University of Kurdistan, P.O.Box 66177-15175, Sanandaj, Iran}
\date{Received: date / Revised version: date}
\maketitle
\begin{abstract}
In the framework of the Heisenberg picture, an alternative derivation of the reduced density matrix of a driven dissipative quantum harmonic oscillator as the prototype of an open quantum system is investigated. The reduced density matrix for different initial states of the combined system is obtained from a general formula, and different limiting cases are studied. Exact expressions for the corresponding characteristic function in quantum thermodynamics and Wigner quasi distribution function are found. A possible generalization based on the Magnus expansion of the evolution operator is presented.
\PACS{
      {05.40.Jc}{Brownian motion}   \and
      {03.65.Yz}{Decoherence; open systems; quantum statistical methods} \and
      {03.67.-a}{Quantum information} \and
      {05.30.-d}{Quantum statistical mechanics}
     } 
\end{abstract}
%
%
\section{Introduction}\label{Introduction}
Experimental techniques in design and manufacturing nanoscale devices to be applied in quantum technology have considerably improved and reached a high level of accuracy in recent years. These devices due to working in the quantum domain are so sensitive to external sources and noises that a precise theoretical understanding of their function is vital to controlling and correcting unwanted behaviors. These devices or systems belong to a wider class of quantum systems interacting with their environment known as open quantum systems \cite{Breuer}. The subject of open quantum systems covers a wide range of applications in quantum physics due to the fact that no quantum system can be completely isolated from its environment. An important paradigm of an open quantum system is the quantum Brownian motion \cite{A1,A2} which has been investigated extensively by different approaches \cite{B1,B2,B3,B4,B5} and appears in miscellaneous problems in physics and chemistry \cite{D1,D2,D3,D4,D5,D6,D7,D8,D9,D10,D11,D12}. Generally, the Hamiltonian of an open quantum system consists of three parts, Hamiltonian of the main system $H_S$, Hamiltonian of the environment $H_R$ and the interaction Hamiltonian $H_{SR}$. The main ingredient in the context of open quantum system theory is the reduced density operator $\roh_S (t)$ describing the main system under consideration. This operator is obtained by tracing out the environmental degrees of freedom of total density matrix $\roh(t)$ describing the combined system. Knowing the explicit form of the reduced density matrix, a complete description of the quantum dynamics of the main subsystem is achievable.

Another issue that should be taken into account while investigating the thermodynamical properties of nanoscale devices working at the quantum regime is the significance of quantum fluctuations that may lead to a reconsideration of thermodynamical laws. Therefore, a precise investigation of the relation between thermodynamics and quantum mechanics is unavoidable and has opened a new issue nowadays referred to as quantum thermodynamics \cite{C1,C2,C3,C4,C5,C6,C7,C8,C9,C10}.

Our aim in the present letter is to introduce an alternative derivation of the reduced density matrix of a driven dissipative quantum harmonic oscillator. Although, the method applied in the present work is based on the existence of exact expressions for the time-evolution of dynamical observables, as a generalization, a perturbative method based on the Magnus expansion (Sec. (\ref{XI})) can also be developed leading to approximate expressions for the time-evolution of the main dynamical variables. The main result of this investigation is the general formula Eq. (\ref{III-8}) and its particular form Eq. (\ref{III-11}) from which exact results can be extracted for different initial states of the combined system or for different limiting cases. Due to the important role played by the characteristic function in quantum thermodynamics, an exact expression for this function is given in Eq. (\ref{CF4}) and its limiting cases are also considered. In the following, the exact Wigner quasi distribution function corresponding to the reduced density matrix is found.
\section{Basics}\label{I}
Here the complex conjugation of an arbitrary c-valued quantity $z$ is denoted by $\bar{z}$, its complex norm by $|z|$ and Laplace transform of an arbitrary function $\varphi(t)$ is denoted by $\mathcal{L}[\varphi(t)]=\tilde{\varphi}(s)$, $(\varphi(t)= \mathcal{L}^{-1}[\tilde{\varphi}(s)]$). The Hamiltonian that we have considered here is the Hamiltonian of a dissipative quantum harmonic oscillator under the influence of a classical external source given by
\bea\label{II-1}
\hat{H} &=& \overbrace{\hbar\omega_0\,\ahd \ah}^{\hat{H}_S}+\overbrace{\sum_j \hbar\oj\bjd \bj}^{\hat{H}_R} + \overbrace{\sum_j[\hbar f_j \bj\,\ahd+\hbar\bar{f}_j \bjd\ah]}^{\hat{H}_{SR}}\nonumber\\
&& +\, \underbrace{\hbar\,K(t)\,\ahd+\hbar\,\bar{K}(t)\,\ah}_{\hat{H}_{ext}},
\eea
where $f_j$'s are coupling constants coupling the oscillator to its environment and $K(t))$ is an arbitrary time-dependent classical external source. Our aim is to find the exact reduced density matrix of the oscillator as the main subsystem in an alternative simple and efficient way. For this purpose, we first find the time-evolution of the oscillator ladder operators in the Heisenberg picture as (App. \ref{A})
\bea\label{II-2}
 \ah (t) &=& G(t)\,\ah(0)-i\zeta (t)-i\sum_j M_j (t)\,\bj(0),\nonumber\\
 \ahd (t) &=& \bar{G}(t)\,\ah(0)+i\bar{\zeta} (t)+i\sum_j \bar{M}_j (t)\,\bjd(0),
\eea
where we have defined
\bea\label{II-3}
&& G(t) = \mathcal{L}^{-1}\bigg[\frac{1}{s+\tilde{\chi}(s)+i\omega_0}\bigg],\nonumber\\
&& \tilde{\chi}(s) = \sum_j \frac{|f_j|^2}{s+i\oj},\nonumber\\
&& M_j (t) = f_j\,\int_0^t dt'\,e^{-i\oj(t-t')}\,G(t'),\nonumber\\
&& \zeta(t) = \int_0^t dt'\,G(t-t')\,K(t').
\eea
From Eqs. (\ref{II-2}) and $[\ah(t),\ahd(t)]=1$ we deduce
\be\label{prob}
|G(t)|^2+\sum_j |M_j (t)|^2=1.
\ee
From Heisenberg equations for reservoir operators we find (App. \ref{A})
\bea\label{bbdager}
\bj (t) &=& \sum_k \Lambda_{jk} (t) \,\hat{b}_k (0)+\Gamma_j (t) \,\hat{a}(0)+\Omega_j (t),\nonumber\\
\bjd (t) &=& \sum_k \bar{\Lambda}_{jk} (t) \,\hat{b}^\dag_k (0)+\bar{\Gamma_j} (t) \,\hat{a}^\dag(0)+\bar{\Omega_j} (t),
\eea
where
\bea\label{bbdefs}
  \Lambda_{jk} (t) &=& e^{-i\oj t}\,\delta_{jk}-\bar{f}_j \int_0^t dt'\,e^{-i\oj (t-t')}\,M_k (t'),\nonumber\\
  \Gamma_j (t) &=& -i\bar{f}_j \int_0^t dt'\,e^{-i\oj (t-t')}\,G(t'),\nonumber\\
  \Omega_j (t) &=& -\bar{f}_j \int_0^t dt'\,e^{-i\oj (t-t')}\,\zeta(t').
\eea
From $[\bj (t),\bjd (t)]=1$ one easily finds
\be
(\Lambda (t)\Lambda^\dag (t))_{jj}+|\Gamma_j (t)|^2=1.
\ee
\section{Matrix elements of the reduced density matrix}\label{II}
Let us assume that the initial density matrix ($\roh(t=0)$), of the oscillator and its environment is a separable state $\roh(0)=\roh_S (0)\otimes\roh_R (0)$. Then, the total density matrix at an arbitrary time $t$ is
\be\label{III-1}
\roh(t)=\Uh (t)\,\roh(0)\,\Uhd (t).
\ee
where the unitary operator $\Uh(t)$ is the evolution operator of the combined system. The reduced density matrix of the oscillator can be obtained by tracing out the environment degrees of freedom of $\roh(t)$ as $\roh_s (t)=\mbox{Tr}_R\{\roh(t)\}$. For the matrix elements of the reduced density matrix in the basis of number states we have
\bea\label{III-2}
\la n|\roh_S (t)|m\ra &=& \la n|\mbox{Tr}_R\{\Uh (t)\,\roh(0)\,\Uhd (t)|m\ra,\nonumber\\
&=& \mbox{Tr}_S \Big(|m\ra\la n|\mbox{Tr}_R \big(\Uh (t)\,\roh(0)\,\Uhd (t)\big)\Big) ,\nonumber\\
&=& \mbox{Tr}\Big\{\big(|m\ra\la n|\otimes I_R\big)\,\Uh (t)\,\roh(0)\,\Uhd (t)\Big\},\nonumber\\
&=& \mbox{Tr}\Big\{\underbrace{\Uhd (t)\big(|m\ra\la n|\otimes I_R\big)\,\Uh (t)}_{\hat{Q}_{mn} (t)}\,\roh(0)\Big\},
\eea
where $\mbox{Tr}_S$ denotes trace over oscillator degrees of freedom, $\mbox{Tr}$ denotes the total trace and $I_R$ is identity operator over the environment Hilbert space. In Eq. (\ref{III-2}) we have defined
\bea\label{Qmn}
\hat{Q}_{mn} (t)=\Uhd (t)\big(|m\ra\la n|\otimes I_R\big)\,\Uh (t),
\eea
with the following properties (App. \ref{B})
\bea\label{Qmn2}
&& \sum_{n=0}^\infty \hat{Q}_{nn}(t)=I,\nn\\
&& \sum_{n=0}^\infty n^s\,\hat{Q}_{nn}(t)=[\ahd(t)\ah(t)]^s.
\eea
Eq. (\ref{Qmn}) can be rewritten in terms of the Heisenberg operators as (App. \ref{C})
\be\label{III-3}
\hat{Q}_{mn} (t)=\frac{1}{\sqrt{m!n!}}\,\sum_{s=0}^\infty \frac{(-1)^s}{s!}\,[\ahd (t)]^{s+m}\,[\ah (t)]^{s+n},
\ee
and plays a fundamental role in what follows. Now we can rewrite Eq. (\ref{III-2}) as
\bea\label{III-4}
\la n|\roh_S (t)|m\ra &=& \mbox{Tr}\{\hat{Q}_{mn} (t)\,\roh (0)\}\nonumber\\
&=& \frac{1}{\sqrt{m!n!}}\,\sum_{s=0}^\infty \frac{(-1)^s}{s!}\,\mbox{Tr}\Big\{[\ahd (t)]^{s+m}\,[\ah (t)]^{s+n}\,\roh_S (0)\otimes \roh_R (0)\Big\},
\eea
on the other hand, from Eq. (\ref{II-2}) we have
\bea\label{III-5}
[\ah (t)]^{s+n} &=& \Big(G(t)\,\ah (0)-i(\zeta(t)+\hat{B})\Big)^{s+n},\nonumber\\
&=& \sum_{r=0}^{s+n}\binom{s+n}{r}[G(t)]^{s+n-r}\,(-i)^r\,[\ah(0)]^{s+n-r}\nonumber\\
&& \times\,\sum_{v=0}^{r}\binom{r}{v}[\zeta (t)]^{r-v}\,[\hat{B}]^{v},
\eea
and similarly
\bea\label{III-6}
[\ahd (t)]^{s+m} &=& \Big(\bar{G}(t)\,\ahd (0)+i(\bar{\zeta}(t)+\hat{B}^\dag)\Big)^{s+m},\nonumber\\
&=& \sum_{r=0}^{s+m}\binom{s+m}{r}[\bar{G}(t)]^{s+m-r}\,(i)^r\,[\ahd(0)]^{s+m-r}\nonumber\\
&& \times\,\sum_{v=0}^{r}\binom{r}{v}[\bar{\zeta} (t)]^{r-v}\,[\hat{B}^\dag]^{v},
\eea
where for convenience we defined
\bea\label{III-7}
\hat{B} &=& \sum_j M_j (t)\,\bj (0),\nonumber\\
\hat{B}^\dag &=& \sum_j \bar{M}_j (t)\,\bjd (0).
\eea
By inserting Eqs. (\ref{III-5}) and (\ref{III-6}) into Eq. (\ref{III-4}), we find
\bea\label{III-8}
\la n|\roh_S (t)|m\ra &=& \frac{1}{\sqrt{n!m!}}\,\sum_{s=0}^\infty \frac{(-1)^s}{s!}\sum_{p=0}^{s+m}\sum_{u=0}^{p}\sum_{r=0}^{s+n}\sum_{v=0}^{r}\nonumber\\
&& \times\binom{s+m}{p}\binom{p}{u}\binom{s+n}{r}\binom{r}{v}\,(i)^{p-r}\nonumber\\
&& \times[\bar{G}(t)]^{s+m-p}[G(t)]^{s+n-r}[\bar{\zeta} (t)]^{p-u}[\zeta (t)]^{r-v}\nonumber\\
&& \times\mbox{Tr}_S \Big([\ahd (0)]^{s+m-p}[\ah (0)]^{s+n-r}\,\roh_S (0)\Big)\nonumber\\
&& \times\mbox{Tr}_R \Big([{\hat{B}}^\dag]^{u}[\hat{B}]^{v}\,\roh_R (0)\Big).
\eea
Eq. (\ref{III-8}) is the most general formula representing the reduced density matrix in the basis of number states. In order to proceed, we assume that the initial state of the environment is a thermal Maxwell-Boltzmann state defined by
\bea\label{MB}
&& \roh_R (0)=\frac{1}{Z_R}\prod_j \otimes \,e^{-\beta\hbar\oj\bjd \bj},\nn\\
&& Z_R=\prod_j z_j,\nn\\
&& z_j=\mbox{Tr}_j [e^{-\beta\hbar\oj\bjd \bj}].
\eea
where $\beta=1/k_B T$, $k_B$ is the Boltzmann constant, $T$ is the temperature of the environment and $\mbox{Tr}_j$ is the trace operator in the Hilbert space of the $j$th oscillator in the environment. By making use of the generating function method it can be proved that (App. \ref{D})
\be\label{III_9}
\mbox{Tr}_R \Big([{\hat{B}}^\dag]^{u}[\hat{B}]^{v}\,\roh_R (0)\Big)=\delta_{uv}\,u!\,[\eta(t)]^u,
\ee
where we have defined
\be\label{III-10}
\eta(t)=\sum_j \frac{|M_j (t)|^2}{e^{\beta\hbar\oj}-1}.
\ee
Therefore,
\bea\label{III-11}
\la n|\roh_S (t)|m\ra &=& \frac{1}{\sqrt{n!m!}}\,\sum_{s=0}^\infty \frac{(-1)^s}{s!}\sum_{p=0}^{s+m}\sum_{r=0}^{s+n}\sum_{u=0}^{\min(p,r)}\nonumber\\
&& \times\,\binom{s+m}{p}\binom{s+n}{r}\binom{p}{u}\binom{r}{u}u!\,[\eta(t)]^u\,(i)^{p-r}\nonumber\\
&& \times\,[\bar{G}(t)]^{s+m-p}[G(t)]^{s+n-r}[\bar{\zeta}(t)]^{p-u}[\zeta (t)]^{r-u}\nonumber\\
&& \times\,\mbox{Tr}_S \Big([\ahd (0)]^{s+m-p}[\ah (0)]^{s+n-r}\,\roh_S (0)\Big).
\eea
Eq. (\ref{III-11}) is the main result of this section. In the last line of this equation, the trace operator acts on subsystem operators at initial time $t=0$ that can be achieved straightforwardly. In the rest of this letter, we will extract different physical results from Eq. (\ref{III-11}) by considering different initial states or conditions on the total Hamiltonian given in Eq. (\ref{II-1}).
\section{The oscillator is initially in a coherent state}\label{III}
In this case we have $\roh_s (0)=|\alpha\ra\la \alpha|$, therefore,
\bea\label{A1}
 \la n|\roh_S (t)|m\ra &=& \frac{1}{\sqrt{n!m!}}\,\sum_{s=0}^\infty \frac{(-1)^s}{s!}\sum_{p=0}^{s+m}\sum_{r=0}^{s+n}\sum_{u=0}^{\min(p,r)}\nonumber\\
&& \times\binom{s+m}{p}\binom{s+n}{r}\binom{p}{u}\binom{r}{u}u!\,[\eta(t)]^u\,(i)^{p-r}\nonumber\\
&& \times[\bar{G}(t)]^{s+m-p}[G(t)]^{s+n-r}[\bar{\zeta} (t)]^{p-u}[\zeta (t)]^{r-u}\nonumber\\
&& \times(\bar{\alpha})^{s+m-p} (\alpha)^{s+n-r},
\eea
which is an exact expression suitable for both analytical and numerical calculations. Let us consider some limiting cases. In low temperature limit ($\beta\rightarrow\infty$), using Eq. (\ref{III-10}) we have $ \eta(t)\rightarrow 0$, therefore, by setting $\eta(t)=0$ in Eq.(\ref{A1}) we obtain
\bea\label{A2}
\la n|\roh_S (t)|m\ra &=& \frac{1}{\sqrt{n!m!}}\,(\alpha G(t)-i\zeta(t))^n (\bar{\alpha} \bar{G}(t)+i\bar{\zeta}(t))^m\,e^{-|\alpha G(t)-i\zeta(t)|^2}.
\eea
By setting $n=m$ in Eq. (\ref{A2}) we find the probability of finding the system in number state $|n\ra$
\bea\label{A3}
\la n|\roh_S (t)|n\ra &=& \frac{|\alpha G(t)-i\zeta(t)|^{2n}}{n!} \,e^{-|\alpha G(t)-i\zeta(t)|^2},
\eea
which is a Poisson distribution with mean number parameter
\be
\la n\ra=|\alpha G(t)-i\zeta(t)|^2.
\ee
\section{The oscillator is initially in a number state}\label{IV}
In this case we have $\roh_s (0)=|k\ra\la k|$, since $\roh_s (t)$ is a hermitian operator we have $\la n|\roh_S (t)|m\ra=\overline{\la m|\roh_S (t)|n\ra}$, so with no lose of generality we can assume $m\geq n$ and find (App. \ref{E})
\bea\label{A4}
\la n|\roh_S (t)|m\ra &=&
\frac{(i\bar{\zeta})^{m-n}}{\sqrt{n!m!}}\sum_{s=0}^\infty \frac{(-1)^s}{s!}\sum_{r=r_{min}}^{s+n}\frac{k!}{(k-s-n+r)!}\nn\\
& \times& \binom{s+m}{m-n+r}\binom{s+n}{r}|G(t)|^{2(s+n-r)}|\zeta|^{2r}\nn\\
& \times& \sum_{u=0}^r\binom{m-n+r}{u}\binom{r}{u}\,u!\,\bigg[\frac{\eta(t)}{|\zeta(t)|^2}\bigg]^u,
\eea
where $r_{min}=\max(s+n-k,0)$. By setting $n=m$, we can find the probability of finding the oscillator in state $|n\ra$ at time $t$ knowing that it was initially prepared in the state $|k\ra$
\bea\label{Pnz}
P^{\zeta\neq 0,T\neq 0}_{k\rightarrow n} (t) &=& \frac{1}{n!}\sum_{s=0}^\infty \frac{(-1)^s}{s!}\sum_{r=r_{min}}^{s+n}\frac{k!}{(k-s-n+r)!}\nn\\
&& \times \binom{s+n}{r}^2 |G(t)|^{2(s+n-r)} |\zeta(t)|^{2r}\nn\\
&& \times \sum_{u=0}^r\binom{r}{u}^2 \,u!\,\bigg[\frac{\eta(t)}{|\zeta(t)|^2}\bigg]^u.
\eea
In zero temperature limit, Eq. (\ref{Pnz}) reduces to
\bea\label{A5}
P^{\zeta\neq 0,T=0}_{k\rightarrow n} (t) &=& \frac{|G(t)|^{2n}}{n!}\sum_{s=0}^\infty \frac{(-1)^s}{s!}|G(t)|^{2s}\nn\\
&& \times\,\sum_{r=r_{min}}^{s+n}\binom{s+n}{r}^2\,\frac{k!}{(k-s-n+r)!}\,\Big |\frac{\zeta(t)}{G(t)}\Big |^{2r}.
\eea
For $k=0$, the initial state is the vacuum state $|0\ra$ that is a coherent state, in this case $r_{min}=s+n$ and Eq. (\ref{A5}) reduces to
\bea\label{A6}
P^{\zeta\neq 0,T=0}_{0\rightarrow n} (t)=\frac{|\zeta|^{2n}}{n!}\,e^{-|\zeta|^2},
\eea
which is a Poisson distribution that could also be found from Eq.(\ref{A3}) by setting $\alpha=0$. Note that in large-time limit $|G(t)|\rightarrow 0$ and the non-vanishing terms in Eq. (\ref{A5}) are obtained by setting $r=s+n$ leading to the same equation Eq. (\ref{A6}) due to the fact that in large-time limit oscillator will decay to its vacuum state.
\section{Expectation values}\label{V}
Let $F(\ahd(0),\ah(0))$ be an arbitrary function in terms of the oscillator ladder operators at the initial time $t=0$. The expectation value $\la F\ra_t$ at arbitrary time $t$ is (App. \ref{F})
\bea\label{expec}
\la F(\ahd(0),\ah(0))\ra_t &=& \mbox{Tr}_S [F(\ahd(0),\ah(0))\,\roh_S (t)],\nn\\
&=& \mbox{Tr} [F(\ahd(t),\ah(t))\,\roh (0)].
\eea
As an example, let $F=\hbar\omega_0 (\ahd\ah+\ah\ahd)/2$ then from Eqs. (\ref{II-2}) and the initial state
\bea
\hat{\rho}(0)=|n\ra\la n|\otimes \frac{e^{-\beta\hat{H}_R (0)}}{Z_R},
\eea
we find the energy of the oscillator at time $t$ as
\bea\label{ener}
\la E\ra_t &=& \hbar\omega_0\,|G(t)|^2 (n+1/2)+|\zeta(t)|^2+\sum_J |M_j (t)|^2\coth(\beta\hbar\omega_j/2),
\eea
leading to a probabilistic interpretation of Eq. (\ref{prob}).
\section{Special limiting cases}
\subsection{The dissipation can be ignored and coupling to the external source is strong}\label{VI}
In this case the coupling constants $f_j$ are zero and from Eqs. (\ref{II-3}) we have
\bea\label{zeroEn}
&& \tilde{\chi}(s)=0\rightarrow G(t)=e^{-i\omega_0 t},\nonumber\\
&& M_j (t)=0 \rightarrow \eta(t)=0.\nn\\
&& \zeta(t)=\int_0^t dt'\,e^{-i\omega_0 (t-t')}\,K(t').
\eea
To proceed let the initial state of the oscillator be a coherent state $\roh_S (0)=|\alpha\ra\la \alpha|$, then from Eq. (\ref{III-11}) we have (App. \ref{G})
\bea\label{imp}
 \la n|\roh_S (t)|m\ra &=& \frac{[\alpha\,e^{-i\omega_0 t}-i\zeta(t)]^n\,[\bar{\alpha}\,e^{i\omega_0 t}+i\bar{\zeta}(t)]^m}{\sqrt{n!\, m!}}\nonumber\\
&& \times\,e^{-|\alpha\,e^{-i\omega_0 t}-i\zeta(t)|^2},
\eea
and the diagonal elements are given by
\bea\label{imp2}
 \la n|\roh_S (t)|n\ra &=& \frac{|\alpha\,e^{-i\omega_0 t}-i\zeta(t)|^{2n}}{n!}\,e^{-|\alpha\,e^{-i\omega_0 t}-i\zeta(t)|^2},\nn\\
\eea
which is a Poisson distribution function with mean value
\be
\la n\ra=|\alpha\,e^{-i\omega_0 t}-i\zeta(t)|^2.
\ee
\subsection{The external source is switched off}\label{VII}
Now let the initial state be the number state $\roh_s (0)=|k\ra\la k|$ with zero coherency. In the absence of a driving force ($\zeta(t)=0$), we set $u=r,\,\,n=m$ in Eq. (\ref{III-11}) and find
\bea\label{IV-1}
\la n|\roh_S (t)|m\ra &=& \delta_{n,m}\,\frac{1}{n!}\,\sum_{s=0}^\infty \frac{(-1)^s}{s!}\,|G(t)|^{2(s+n)}\nn\\
&& \times \sum_{r=r_{min}}^{s+n}\frac{k!\,r!\,[\eta(t)/|G(t)|^2]^r}{(k-s-n+r)!}\binom{s+n}{r}^2.\nn\\
\eea
Therefore, the evolved reduced density matrix has remained diagonal with zero coherency and diagonal elements are given by
\bea\label{IV-2}
 P^{\zeta=0,T\neq 0}_{k\rightarrow n} (t) &=& \frac{|G(t)|^{2n}}{n!}\sum_{s=0}^\infty \frac{(-1)^s |G(t)|^{2s}}{s!}\nonumber\\
&& \times\sum_{r=r_{min}}^{s+n}{\binom{s+n}{r}}^2 \frac{k!\,r!}{(k-s-n+r)!}\bigg[\frac{\eta(t)}{|G(t)|^2}\bigg]^r.
\eea
In low temperature limit, we set $r=0$ in Eq. (\ref{IV-2}) and find
\bea\label{IV-3}
 P^{\zeta=0,T=0}_{k\rightarrow n} (t)=\theta(k-n)\,\binom{k}{n}\Big(|G(t)|^2\Big)^n\,\Big(1-|G(t)|^2\Big)^{k-n},
\eea
which is a binomial distribution with parameter $|G(t)|^2$. The Heaviside step function in Eq. (\ref{IV-3}) is defined by
\bea
\theta(k-n)=\left\{
              \begin{array}{ll}
                1, & k\geq n \\
                0, & k< n
              \end{array}
            \right.
\eea
and shows that only transitions to lower energy levels ($k\geq n$) are possible.

Now let the initial state be a coherent state $\roh_s (0)=|\alpha\ra\la \alpha|$, and the preferred basis be the number states, then in the absence of external source ($\zeta(t)=0$), we set $p=r=u$ in Eq. (\ref{A1}) and with no loss of generality we can assume $m\geq n$, therefore,
\bea\label{IV-4}
\la n|\roh_S (t)|m\ra &=& \frac{1}{\sqrt{n!m!}}\,\sum_{s=0}^\infty \frac{(-1)^s}{s!}[\bar{\alpha}\bar{G}(t)]^{s+m}[\alpha G(t)]^{s+n}\nn\\
&& \times \sum_{r=0}^{s+n}\binom{s+m}{r}\binom{s+n}{r}\,r!\,\bigg[\frac{\eta(t)}{|G(t)|^2}\bigg]^r.
\eea
From Eq. (\ref{IV-4}) it is seen that the coherency of the density matrix vanishes at large-time limit since the non-diagonal elements tend to zero in this limit (App. \ref{H}).

In low-temperature limit Eq. (\ref{IV-4}) reduces to
\bea\label{IV-5}
&& \la n|\roh_S (t)|m\ra =\frac{[\bar{\alpha}\bar{G}(t)]^{m}[\alpha G(t)]^{n}}{\sqrt{n! m!}}\,e^{-|\alpha G(t)|^2},
\eea
and the probability of finding the oscillator in the number state $|n\ra$ is a Poisson distribution
\bea\label{IV-6}
&& \la n|\roh_S (t)|n\ra =\frac{|\alpha\,G(t)|^{2n}}{n!}\,e^{-|\alpha G(t)|^2}.
\eea
\section{Quantum thermodynamics. The Characteristic function}\label{VIII}
Let $P(Q,t)$ be the probability distribution for the heat amount $Q$ to be transferred to the environment between times $t=0$ and $t$. Then \cite{Talkner,Esposito}
\bea\label{CF1}
P(Q,t) &=& \sum_{e_1,e_2} \delta(e_2-e_1-Q)\,P[e_1\rightarrow e_2;t]\,P[e_1],
\eea
where $P[e_1]$ is the probability of obtaining $e_1$ when measuring environment energy $\hat{H}_R$ at $t=0$ and $P[e_1\rightarrow e_2;t]$ is the conditional probability that a measurement of $\hat{H}_R$ gives $e_2$ at time $t$ when it gave $e_1$ at $t=0$. The characteristic function is defined by the Fourier transform
\bea\label{CF2}
G(\nu,t) &=& \int_{-\infty}^{\infty} dQ\,P(Q,t)e^{i Q\nu},\nonumber\\
         &=& \sum_{e_1,e_2} P[e_1\rightarrow e_2;t]\,P[e_1]\,e^{i\nu (e_2-e_1)}.
\eea
Let the initial density matrix of the total system be a factorized state $\rho(0)=\rho_S(0)\otimes \rho_R (0)$, where $\rho_S(0)$ is the initial density matrix of the oscillator and $\rho_R (0)$ is the initial density matrix of the reservoir which we assume to be a thermalized state
\bea\label{CF3}
\rho_R (0) &=& e^{-\beta\hat{H}_R}/Z_R,\nonumber\\
Z_R &=& \mbox{Tr}[e^{-\beta\hat{H}_R}].
\eea
Now we have
\bea\label{CF4}
P[e_1\rightarrow e_2;t]\,P[e_1] &=& \la e_2|\hat{U}(t) \bigg(\rho_S (0)\otimes |e_1\ra\la e_1|\bigg) \hat{U}^\dag (t) |e_2\ra\,\la e_1|\rho_R (0)|e_1\ra,
\eea
therefore,
\bea
G(\nu,t) &=& \sum_{e_1,e_2}\la e_2|\mbox{Tr}_{S}\bigg[\hat{U}(t) \Big(\rho_S (0)\otimes |e_1\ra\la e_1|\Big) \hat{U}^\dag (t)\bigg] |e_2\ra\,\la e_1|\rho_R (0)|e_1\ra\,e^{i\nu (e_2-e_1)},\nonumber\\
         &=& \sum_{e_2} \la e_2|\mbox{Tr}_{S}\bigg[ \hat{U}(t) \bigg(e^{-i\nu\hat{H}_R}\rho_S (0)\otimes\Big[\sum_{e_1} \la e_1|\rho_R (0)|e_1\ra |e_1\ra\la e_1|\Big]\bigg)\hat{U}^{\dag} (t)\bigg]\nonumber\\
&&\times\,e^{i\nu\hat{H}_R} |e_2\ra,\nonumber\\
         &=& \mbox{Tr}_R \mbox{Tr}_S \bigg[\hat{U} (t) e^{-i\nu\hat{H}_R}\rho_S(0)\otimes \rho_R (0)\,\hat{U}^\dag (t) e^{i\nu\hat{H}_R}\bigg],\nonumber\\
         &=& \mbox{Tr} \bigg[\hat{U}^\dag (t) e^{i\nu\hat{H}_R}\hat{U} (t) e^{-i\nu\hat{H}_R}\,\rho(0)\bigg],\label{CF5}\nonumber\\
         &=& \mbox{Tr} \bigg[e^{i\nu\hat{H}_R (t)}e^{-i\nu\hat{H}_R}\,\rho(0)\bigg].
\eea
In Eq. (\ref{CF5}), $\mbox{TR}_{S(R)}$ means taking trace over system(reservoir) degrees of freedom and $\hat{H}_R (t)$ is the reservoir Hamiltonian in Heisenberg picture defined by
\bea\label{CF6}
\hat{H}_R (t) &=& \hat{U}^\dag (t) \hat{H}_R \hat{U} (t)=\sum_j \hbar\oj\bjd (t) \bj (t),
\eea
where $\bj (t)$ and $\bjd (t)$ are given by Eq. (\ref{bbdager}). From the characteristic function $G(\nu,t)$, the moments of $Q$ can be found as
\bea\label{CF7}
\la Q^n (t) \ra=(-i)^n\,\frac{d^n G(\nu,t)}{d\, \nu^n}\bigg |_{\nu=0}.
\eea
As an example, let the oscillator be initially prepared in the number state $\rho_S (0)=|n\ra\la n|$, using Eq. (\ref{CF5}), the average heat transferred to the reservoir is
\bea\label{CF8}
\la Q (t) \ra &=& (-i)\,\frac{d G(\nu,t)}{d\, \nu}\bigg |_{\nu=0},\nonumber\\
              &=& -i\mbox{Tr}\bigg[i\hat{H}_R (t)e^{i\nu\hat{H}_R (t)}e^{-i\nu\hat{H}_R (0)}+e^{i\nu\hat{H}_R (t)}(-i\hat{H}_R (0))e^{-i\nu\hat{H}_R (0)}\bigg]_{\nu=0},\nonumber\\
              &=& \mbox{Tr}\Big[\big(\hat{H}_R (t)-\hat{H}_R (0)\big)\,\rho (0)\Big],\nonumber\\
              &=& \sum_j \hbar\oj\,\mbox{Tr}\bigg[\Big(\bjd (t) \bj (t)-\bjd (0) \bj (0)\Big) \rho_S (0)\otimes\rho_R (0)\bigg].
\eea
By inserting Eqs. (\ref{bbdager}) into Eq. (\ref{CF8}) we find
\bea\label{CF9}
\la Q (t) \ra &=& n\,\sum_j \hbar\oj |\Gamma_j (t)|^2+\sum_j \hbar\oj |\Omega_j (t)|^2\nonumber\\
              && + \sum_{j,k} \frac{\hbar\oj |\Lambda_{jk} (t)|^2}{e^{\beta\hbar\omega_k}-1}.
\eea
In zero temperature ($\beta\rightarrow\infty$) and in the absence of external source ($\Omega_j (t)=0$), we have
\be
\la Q (t) \ra=n\,\sum_j \hbar\oj |\Gamma_j (t)|^2.
\ee
\section{Thermal equilibrium}\label{IX}
In the absence of an external source ($\zeta=0$) the oscillator tends to an equilibrium state in large-time limit for arbitrary initial state. This can be easily deduced by setting $r=p=u,\,\,(m\geq n)$ in Eq. (\ref{III-11}), we have
\bea\label{T1}
 \la n|\roh_S (t)|m\ra &=& \frac{1}{\sqrt{n!m!}}\sum_{s=0}^\infty \frac{(-1)^s}{s!}\sum_{r=0}^{s+n}\binom{s+m}{r}\binom{s+n}{r}\nonumber\\
&& \times\,[\bar{G}(t)]^{s+m-r}[G(t)]^{s+n-r}\,r!\,[\eta(t)]^r\nonumber\\
&& \times\,\mbox{Tr}_S \Big([\ahd (0)]^{s+m-r}[\ah (0)]^{s+n-r}\,\roh_S (0)\Big),
\eea
in large-time limit we have $|G(t)|\rightarrow 0$, leading to (App. \ref{I})
\bea\label{T2}
&& \la n|\roh_S (t)|m\ra = \delta_{n,m}\,\frac{[\eta(t)]^n}{[1+\eta(t)]^{n+1}},
\eea
which is a thermal state with mean number $\la n\ra=\eta(t)$.
\section{Wigner function}\label{X}
An important quasi distribution function on phase space is the Wigner function. In this section we find an expression for the Wigner function corresponding to a driven dissipative harmonic oscillator. The components of the reduced density matrix in the continuous position basis $\{|x\ra\}$ are
\bea\label{W1}
\la x|\roh_S (t)|x'\ra &=& \sum_{n,m=0}^\infty \la x|n\ra\la n|\roh_S (t)|m\ra\la m|x'\ra,\nonumber\\
&=& \sum_{n,m=0}^\infty \psi_n (x)\bar{\psi}_m (x')\,\la n|\roh_S (t)|m\ra,
\eea
where $\psi_n (x)$ is the $n$th eigenvector of the free oscillator Hamiltonian. The Wigner quasi distribution function is defined by
\bea\label{W2}
W(x,p;t) &=& \frac{1}{\pi\hbar}\int_{-\infty}^\infty dy\,\la x+y|\roh_S (t)|x-y\ra\,e^{\frac{-2i}{\hbar}py},\nn\\
&=& \sum_{n,m=0}^\infty \la n|\roh_S (t)|m\ra\,\bigg\{\frac{1}{\pi\hbar}\int\limits_{-\infty}^\infty dy\,\psi_n (x+y)\,\bar{\psi}_m (x-y)e^{\frac{-2ipy}{\hbar}}\bigg\}.
\eea
As a special case, let the initial state of the oscillator be the number state $\roh_S (0)=|k\ra\la k|$, and the external source be switched off, then
\bea\label{W3}
&& W(x,p;t)=\sum_{n=0}^\infty\,P^{\zeta=o, T\neq 0}_{k\rightarrow n} (t)\,W_n (x,p;t),
\eea
where
\bea\label{W4}
W_n (x,p;t)=\frac{1}{\pi\hbar}\int\limits_{-\infty}^\infty dy\,\psi_n (x+y)\bar{\psi}_n (x-y)e^{\frac{-2ipy}{\hbar}},
\eea
is the Wigner function corresponding to the pure state $|n\ra\la n|$ and $P^{\zeta=o, T\neq 0}_{k\rightarrow n} (t)$ is given by Eq. (\ref{IV-2}). From Eq. (\ref{W3}) it is seen that the quasi distribution function $W$ is the average of quasi distributions $W_n$ with respect to the probability distribution $P^{\zeta=o, T\neq 0}_{k\rightarrow n} (t)$. In low temperature limit, using Eq. (\ref{IV-3}), we find
\bea\label{W5}
W(x,p;t) &=& \sum_{n=0}^k \binom{k}{n}\Big(|G(t)|^2\Big)^n\Big(1-|G(t)|^2\Big)^{k-n}\,W_n (x,p;t),\nn\\
&=& \la W_n (x,p;t) \ra_{binomial}.
\eea
As another case, let the initial state of the oscillator be a coherent state $\hat{\rho}_S (0)=|\alpha\ra\la \alpha|$, then in the zero temperature ($\beta\rightarrow\infty$) we have from Eq. (\ref{A2})
\bea
W(x,p;t) &=& \frac{1}{\hbar\pi}\int_{-\infty}^{\infty} dy\,\la x+y|\sum_{n=0}^\infty \frac{(\alpha\,G-i \zeta)^n}{\sqrt{n!}}|n\ra \sum_{m=0}^\infty \la m|\frac{(\bar{\alpha}\,\bar{G}+i \bar{\zeta})^m}{\sqrt{m!}}|x-y\ra\,e^{\frac{-2ipy}{\hbar}}\nonumber\\
&& \times\,e^{-|\alpha\,G-i \zeta|^2},\nonumber\\
         &=& \frac{1}{\hbar\pi}\int_{-\infty}^{\infty} dy\,\la x+y|\alpha\,G(t)-i \zeta (t)\ra\la \alpha\,G(t)-i \zeta (t)|x-y\ra\,e^{\frac{-2ipy}{\hbar}},
\eea
which is the Wigner function corresponding to the following evolved coherent state
\be
\hat{\rho}_S (t)=|\alpha\,G(t)-i \zeta (t)\ra\la \alpha\,G(t)-i \zeta (t)|.
\ee
\section{Generalisation}\label{XI}
In the present work we have concentrated on the Hamiltonian Eq. (\ref{II-1}) which is the Hamiltonian of a driven dissipative harmonic oscillator. However, there are few systems that their Heisenberg equations are integrable. So a perturbative approach is unavoidable. Among the perturbative methods, the Magnus expansion \cite {Magnus} has some preferences. The main preference is the preservation of the unitarity of the evolution operator at any order of approximation. Having an approximate evolution operator $\hat{U}_{appr.}(t)$ we find an approximation expression for an arbitrary dynamical variable $\hat{A}$ in Heisenberg picture as
\be\label{gen1}
A(t)\approx\hat{U^\dag}_{appr.}(t)\,A\,\hat{U}_{appr.}(t).
\ee
Let the total Hamiltonian be given by
\be\label{gen2}
\hat{H}(t)=\hat{H}_0 +\epsilon (t) \hat{H}_1,
\ee
where $[\hat{H}_0, \hat{H}_1]\neq 0$. The exact evolution operator $\hat{U}(t)$ satisfies the Schr\"{o}dinger equation
\be\label{gen3}
\frac{d\hat{U}(t)}{dt}=-\frac{i}{\hbar}\hat{H}(t)\,\hat{U}(t).
\ee
To find an approximate unitary solution, following \cite{Magnus} we assume
\bea\label{gen4}
\hat{U}(t) &=& e^{\hat{u}(t)},\nonumber\\
\hat{u}(t) &=& \sum_{k=1}^\infty \hat{u}_k (t),
\eea
then
\bea\label{gen5}
\hat{u}_1 (t) &=& -\frac{i}{\hbar}\,\int_0^t dt_1\,\hat{H}(t_1),\nonumber\\
\hat{u}_2 (t) &=& \frac{1}{2}(-\frac{i}{\hbar})^2\,\int_0^tdt_1\,\int_0^{t_1} dt_2\,[\hat{H}(t_1), \hat{H}(t_2)],\nonumber\\
\hat{u}_3 (t) &=& \frac{1}{6}(-\frac{i}{\hbar})^3\,\int_0^tdt_1\,\int_0^{t_1} dt_2\,\int_0^{t_2} dt_3\,\Big\{[\hat{H}(t_1),[\hat{H}(t_2), \hat{H}(t_3)]]\nonumber\\
&& +[\hat{H}(t_3), [\hat{H}(t_2), \hat{H}(t_1)]]\Big\},\nonumber
\eea
therefore,
\be\label{gen6}
\hat{U}_{appr.}(t)\approx e^{\hat{u}_1 +\hat{u}_2 +\hat{u}_3},
\ee
and the time-evolution of an arbitrary dynamical variable $\hat{A}$ in Heisenberg picture is approximately given by
\be\label{gen7}
\hat{A}(t)=\hat{U}^\dag_{appr.}(t)\,\hat{A}(0)\,\hat{U}_{appr.}(t),
\ee
and similar steps can be followed to find an approximate reduced density matrix in a preferred basis.
\section{Conclusions}\label{XII}
An alternative derivation of the reduced density matrix of a driven dissipative quantum harmonic oscillator as the prototype of an open quantum system was introduced in the Heisenberg picture. The reduced density matrix for different initial states of the combined system was obtained from a general formula, and different limiting cases were studied. Exact expressions for the corresponding characteristic function in quantum thermodynamics and Wigner quasi distribution function were found. Though the method introduced here was based on the existence of exact expressions for the time-evolution of dynamical observables, as a generalization, a perturbative method based on the Magnus expansion could be developed leading to approximate expressions for the time-evolution of the main dynamical variables in Heisenberg picture.
\appendix
\section{Derivation of Eqs. (\ref{II-2}) and (\ref{bbdager})}\label{A}
From Hamiltonian Eq. (\ref{II-1}) and Heisenberg equations of motion for $\ah$ and $\bj$ we have
\bea\label{1}
\dot{\ah} &=& \frac{1}{i\hbar}[\ah,\hat{H}]=-i\omega_0\ah-i\sum_j f_j \bj-i K(t).
\eea
\bea\label{2}
\dot{\hat{b}}_j &=& \frac{1}{i\hbar}[\bj,\hat{H}]=-i\oj\bj-i \bar{f}_j \ah.
\eea
The solution of Eq. (\ref{2}) is
\bea\label{3}
\bj (t)=e^{-i\oj t} \bj (0)-i \bar{f}_j \int_0^t dt'\,e^{-i \oj (t-t')} \ah (t'),
\eea
by inserting this solution into Eq. (\ref{1}), we find
\bea\label{4}
\dot{\ah}+i\omega_0 \,\ah+\int_0^t dt'\,\chi(t-t')\,\ah (t') &=& -i\sum_j f_j\,e^{-i \oj t}\,\bj (0)-i\, K(t),
\eea
where the response function of the medium is defined by
\bea\label{5}
\chi(t-t')=\sum_j |f_j|^2\,e^{-i\oj (t-t')}.
\eea
By taking the Laplace transform of both sides of Eq. (\ref{4}) we obtain
\bea\label{6}
\tilde{a}(s) &=& \frac{1}{s+i\omega_0+\tilde{\chi}(s)}\,\ah (0)-i\frac{\tilde{K}(s)}{s+i\omega_0+\tilde{\chi}(s)}\nn\\
&& -i \sum_j \frac{f_j\,\bj (0)}{(s+i\omega_0+\tilde{\chi}(s))(s+i\,\oj)},
\eea
and by taking inverse Laplace transform we finally find
\bea\label{7}
\ah (t)=G(t)\,\ah (0)-i\sum_j M_j (t)\,\bj (0)-i \zeta(t),
\eea
with the hermitian conjugation
\bea
\ah^{\dag} (t)=\bar{G}(t)\,\ah^{\dag} (0)+i\sum_j \bar{M}_j (t)\,\bj^{\dag} (0)+i \bar{\zeta}(t),
\eea
where
\bea\label{8}
&& G(t)=\mathcal{L}^{-1}\bigg[\frac{1}{s+i\omega_0+\tilde{\chi}(s)}\bigg],\nn\\
&& M_j (t)=f_j\,\int_0^t dt'\,e^{-i\oj (t-t')}\,G(t'),\nn\\
&& \zeta(t)=\int_0^t dt'\,G(t-t')\,K(t').
\eea
By inserting Eq. (\ref{7}) into Eq. (\ref{3}) we find
\bea
\bj (t) &=& \sum_k \bigg\{\underbrace{e^{-i \oj t}\,\delta_{jk}-\bar{f}_j \int_0^t dt'\,e^{-i\oj (t-t')}\,M_{k} (t')}_{\Lambda_{jk} (t)}\bigg\}\,\hat{b}_k (0)\nonumber\\
        && \underbrace{-i \bar{f}_j \int_0^t dt'\,e^{-i\oj (t-t')}\,G_{k} (t')}_{\Gamma_j (t)}\,\hat{a} (0)\nonumber\\
        && \underbrace{- \bar{f}_j \int_0^t dt'\,e^{-i\oj (t-t')}\,\zeta_{k} (t')}_{\Omega_j (t)}.
\eea
Therefore,
\bea
\bj (t) &=& \sum_k \Lambda_{jk} (t)\,\hat{b}_k (0)+\Gamma_j (t) \,\hat{a} (0)+\Omega_j (t),\nonumber\\
\bjd (t) &=& \sum_k \bar{\Lambda}_{jk} (t)\,\hat{b}^\dag_k (0)+\bar{\Gamma}_j (t) \,\hat{a}^\dag (0)+\bar{\Omega}_j (t).
\eea
\section{Derivation of Eqs. (\ref{Qmn2})}\label{B}
\bea
\sum_{n=0}^\infty \hat{Q}_{nn} (t) &=& \sum_{n=0}^\infty \hat{U}^{\dag} (t) \,|n\ra\la n|\otimes I_R\,\hat{U} (t),\nn\\
&=& \hat{U}^{\dag} (t) \,I_S\otimes I_R\,\hat{U} (t)=\hat{U}^{\dag} (t)\,\hat{U} (t)=I.\nn\\
\eea
\bea
\sum_{n=0}^\infty n^s\,\hat{Q}_{nn} (t) &=& \sum_{n=0}^\infty n^s\,\hat{U}^{\dag} (t) \,|n\ra\la n|\otimes I_R\,\hat{U} (t),\nn\\
&=& \hat{U}^{\dag} (t)\sum_{n=0}^\infty n^s\,|n\ra\la n|\otimes I_R\,\hat{U} (t),\nn\\
&=& \hat{U}^{\dag} (t) (\ahd (0)\,\ah (0))^s\,\hat{U} (t),\nn\\
&= & (\ahd (t)\,\ah (t))^s.
\eea
\section{Derivation of Eq. (\ref{III-3})}\label{C}
\bea\label{8}
\hat{Q}_{nn} (t) &=& \hat{U}^{\dag} (t) |m\ra\la n|\otimes I_R\,\hat{U} (t),\nn\\
&=& \hat{U}^{\dag} (t)\,\frac{(\ahd (0))^m}{\sqrt{m!}}|0\ra\la 0|\frac{\ah (0))^n}{\sqrt{n!}}\otimes I_R\,\hat{U} (t).
\eea
On the other hand \cite{Louisell}
\be\label{L}
|0\ra\la 0|=\sum_{s=0}^\infty \frac{(-1)^s}{s!}\,(\ahd (0))^s(\ah (0))^s,
\ee
by inserting Eq. (\ref{L}) into Eq. (\ref{8}) we deduce
\bea
\hat{Q}_{nn} (t) &=& \frac{1}{\sqrt{m! n!}}\,\sum_{s=0}^\infty \frac{(-1)^s}{s!}\,\hat{U}^{\dag} (t)\,(\ahd (0))^{m+s}\,(\ah (0))^{n+s}\otimes I_R\,\hat{U} (t),\nn\\
&=& \frac{1}{\sqrt{m! n!}}\,\sum_{s=0}^\infty \frac{(-1)^s}{s!}\,(\ahd (t))^{m+s}(\ah (t))^{n+s}.\nn\\
\eea
\section{Derivation of Eq. (\ref{III_9})}\label{D}
We have
\bea\label{15}
&& \mbox{Tr}_R \Big[(\hat{B}^{\dag})^u (\hat{B})^v\,\hat{\rho}_R (0)\Big]=\bigg(\frac{\partial}{\partial J}\bigg)^u\,\bigg(\frac{\partial}{\partial \bar{J}}\bigg)^v\,\mbox{Tr}_R \Big[e^{J\hat{B}^{\dag}} \,e^{\bar{J}\hat{B}}\,\hat{\rho}_R (0)\Big]\bigg |_{J=\bar{J}=0}.
\eea
By inserting the definitions
\bea
\hat{B}=\sum_k M_k (t)\,\bk (0),\nn\\
\hat{B}^{\dag}=\sum_k \bar{M}_k (t)\,\bkd (0),
\eea
into the generating function defined by Eq. (\ref{15}) we find
\bea\label{Gen}
&& \mbox{Tr}_R \Big[e^{J\hat{B}^{\dag}} \,e^{\bar{J}\hat{B}}\,\hat{\rho}_R (0)\Big]=\prod_k \underbrace{\mbox{Tr}_k \bigg(e^{J \bar{M}_k (t)\bkd (0)}e^{\bar{J} M_k (t) \bk (0)}
\frac{e^{-\beta\hbar\ok\,\bkd\bk}}{z_k}\bigg)}_{I_k},
\eea
where $\mbox{Tr}_k$ means taking trace over the Hilbert space of the kth oscillator in the environment and $z_k$ is the corresponding partition function
\be
z_k=\mbox{Tr}_k\big(e^{-\beta\hbar\ok\,\bkd\bk}\big).
\ee
Now we have
\bea
&& I_k = \frac{1}{z_k}\sum_{n_k=0}^\infty e^{-\beta\hbar\ok n_k}\,\la n_k|e^{J \bar{M}_k (t)\bkd (0)}e^{\bar{J} M_k (t) \bk (0)}|n_k\ra,\nn\\
&& = \frac{1}{z_k}\sum_{n_k=0}^\infty e^{-\beta\hbar\ok n_k}\sum_{l=0}^{n_k} \frac{|J|^{2l}\,|M_k (t)|^{2l}}{(l!)^2}\la n_k |(\bkd)^l (\bk)^l|n_k\ra,\nn\\
&& = \frac{1}{z_k}\sum_{n_k=0}^\infty e^{-\beta\hbar\ok n_k}\sum_{l=0}^{n_k} \frac{|J|^{2l}\,|M_k (t)|^{2l}}{(l!)^2}\,\frac{n_k!}{(n_k-l)!},\nn\\
&& = e^{-J\bar{J}\,\frac{|M_k (t)|^2}{1-e^{\beta\hbar\ok}}}.
\eea
Therefore, the generating function is
\be
\mbox{Tr}_R \Big[e^{J\hat{B}^{\dag}} \,e^{\bar{J}\hat{B}}\,\hat{\rho}_R (0)\Big]=e^{-J\bar{J}\,\frac{|M_k (t)|^2}{1-e^{\beta\hbar\ok}}}=e^{-J\bar{J}\,\eta (t)}.
\ee
By making use of Eq. (\ref{15}) we finally find
\bea
\mbox{Tr}_R \Big[(\hat{B}^{\dag})^u (\hat{B})^v\,\hat{\rho}_R (0)\Big] &=& (\partial/\partial J)^u\,(\partial/\partial \bar{J})^v\,e^{-J\bar{J}\,\eta (t)}\bigg|_{J=\bar{J}=0},\nn\\
 &=& \delta_{u v}\,u!\,[\eta (t)]^u.
\eea
\section{Derivation of Eq. (\ref{A4})}\label{E}
We have
\bea\label{E1}
\la n|\roh_S (t)|m\ra &=& \frac{1}{\sqrt{n!m!}}\,\sum_{s=0}^\infty \frac{(-1)^s}{s!}\sum_{p=0}^{s+m}\sum_{r=0}^{s+n}\sum_{u=0}^{\min(p,r)}\,\binom{s+m}{p}\binom{s+n}{r}\binom{p}{u}\binom{r}{u}u!\,[\eta(t)]^u\nonumber\\
&& \times\,[\bar{G}(t)]^{s+m-p}[G(t)]^{s+n-r}[\bar{\zeta} (t)]^{p-u}[\zeta (t)]^{r-u}\,(i)^{p-r}\nonumber\\
&& \times\,\sqrt{\frac{k!}{(k-s-n+r)!}}\sqrt{\frac{k!}{(k-s-m+p)!}}\,\delta_{m-p,n-r},\nonumber\\
&=&  \frac{i^{m-n}}{\sqrt{n!m!}}\,\sum_{s=0}^\infty \frac{(-1)^s}{s!}\sum_{r=r_{min}}^{s+n}\sum_{u=0}^r\,\binom{s+m}{m-n+r}\binom{s+n}{r}\binom{m-n+r}{u}\binom{r}{u}\nn\\
&& \times\,u!\,[\eta(t)]^u\, \frac{k!}{(k-s-n+r)!}\,|G(t)|^{2(s+n-r)}(\bar{\zeta})^{m-n+r}\,(\zeta)^r\,|\zeta|^{-2u},\nn\\
&=& \frac{(i\bar{\zeta})^{m-n}}{\sqrt{n!m!}}\sum_{s=0}^\infty \frac{(-1)^s}{s!}\sum_{r=r_{min}}^{s+n}\frac{k!}{(k-s-n+r)!}\,\binom{s+m}{m-n+r}\binom{s+n}{r}\nonumber\\
&& \times\,|G(t)|^{2(s+n-r)}|\zeta|^{2r}\,\sum_{u=0}^r\binom{m-n+r}{u}\binom{r}{u}\,u!\,\bigg[\frac{\eta(t)}{|\zeta(t)|^2}\bigg]^u,
\eea
where $r_{min}=\max(s+n-k,0)$.
\section{Derivation of Eq. (\ref{expec})}\label{F}
\bea\label{26}
\la F(\ahd(0),\ah(0))\ra_t &=& \mbox{Tr}_S [F(\ahd(0),\ah(0))\,\roh_S (t)],\nn\\
&=& \mbox{Tr}_S [F(\ahd(0),\ah(0))\,\mbox{Tr}_R\roh (t)],\nn\\
&=& \mbox{Tr} [F(\ahd(0),\ah(0))\,\roh (t)],\nn\\
&=& \mbox{Tr} [F(\ahd(0),\ah(0))\,\hat{U} (t)\,\roh (0)\,\hat{U}^{\dag} (t)],\nn\\
&=& \mbox{Tr} [\hat{U}^{\dag} (t)\,F(\ahd(0),\ah(0))\,\hat{U} (t)\roh (0)],\nn\\
&=& \mbox{Tr} [F(\ahd(t),\ah(t))\,\roh (0)],\nn\\
\eea
\section{Derivation of Eq. (\ref{imp})}\label{G}
In the absence of dissipation ($f_k=0,\,\,\,k=0,1,2,\cdots $), we have from Eq. (\ref{III-11})
\bea
&& \tilde{\chi} (s)=0 \rightarrow G(t)=e^{-i\omega_0 t},\nn\\
&& \eta (t)=0.
\eea
Therefore,
\bea
\la n|\roh_S (t)|m\ra &=& \frac{1}{\sqrt{n! m!}}\sum_{s=0}^\infty \frac{(-1)^s}{s!}\sum_{p=0}^{s+m}\sum_{r=0}^{s+n}\,\binom{s+m}{p}\binom{s+n}{r}\,e^{-i\omega_0 t (s+n-r)}e^{i\omega_0 t (s+m-p)}\nn\\
&& \times \bar{\zeta}^p \,\zeta^r \,\alpha^{s+n-r} \,\bar{\alpha}^{s+m-p}\,(i)^{p-r},\nn\\
&=& \frac{1}{\sqrt{n! m!}}\sum_{s=0}^\infty \frac{(-1)^s}{s!}|\alpha|^{2s}\,\alpha^n\,\bar{\alpha}^m\,e^{-i n \omega_0 t} e^{i m \omega_0 t}\nn\\
&& \times \sum_{p=0}^{s+m}\binom{s+m}{p}e^{-i p \omega_0 t}\,(i\bar{\zeta})^p\,(\bar{\alpha})^{-p}\,\sum_{r=0}^{s+n}\binom{s+n}{r}e^{i r \omega_0 t}\,(-i\zeta)^r\,\alpha^{-r},\nn\\
&=& \frac{1}{\sqrt{n! m!}}\sum_{s=0}^\infty \frac{(-1)^s}{s!} |\alpha|^{2s}\,\alpha^n\,\bar{\alpha}^m\,e^{-i n \omega_0 t} e^{i m \omega_0 t}\nn\\
&& \times \bigg(1+\frac{i\bar{\zeta}e^{-i\omega_0 t}}{\bar{\alpha}}\bigg)^{s+m}\bigg(1-\frac{i\zeta}{\alpha}e^{i\omega_0 t}\bigg)^{s+n},\nn\\
&=& \frac{1}{\sqrt{n! m!}}\sum_{s=0}^\infty \frac{(-1)^s}{s!} |\alpha-i\zeta\,e^{i\omega_0 t}|^{2s}\,(\bar{\alpha} \,e^{i\omega_0 t}+i\bar{\zeta})^m\,(\alpha \,e^{-i\omega_0 t}-i\zeta)^n,\nn\\
&=& \frac{(\alpha \,e^{-i\omega_0 t}-i\zeta)^n(\bar{\alpha} \,e^{i\omega_0 t}+i\bar{\zeta})^m}{\sqrt{n! m!}}\,e^{-|\alpha \,e^{-i\omega_0 t}-i\zeta|^2}.
\eea
\section{Vanishing of coherency in Eq. (\ref{IV-4})}\label{H}
In large-time limit, we have $G(t)\rightarrow 0$ and for the non diagonal matrix elements of the reduced density matrix ($n\neq m$) we easily see that the maximum degree of $G(t)$ in denominator is smaller than its degree in numerator, since
\be
2r=2(s+\min(m,n))<2s+m+n,\,\,\,(m\neq n),
\ee
therefore, the non diagonal elements tend to zero and accordingly there is no coherency in the preferred number basis $|n\ra$.
\section{Derivation of Eq. (\ref{T2})}\label{I}
In large-time limit $G(t)\rightarrow 0$, so by setting $s+n-r=0$ and $s+m-r=0$ we find $n=m$, leading to
\bea
\la n|\roh_S (t)|m\ra &=& \delta_{n,m}\,\frac{1}{\sqrt{n! m!}}\sum_{s=0}^\infty \frac{(-1)^s}{s!}\,(s+n)!\,[\eta(t)]^{s+n},\nn\\
&=& \delta_{n,m}\,\frac{\eta(t)^n}{\sqrt{n! m!}}\,\sum_{s=0}^\infty (-1)^s\,\frac{(s+n)!}{s!}\,[\eta(t)]^{s},\nn\\
&=& \delta_{n,m}\,\frac{\eta(t)^n}{n!}\,\frac{n!}{[1+\eta(t)]^{n+1}}\nn\\
&=& \delta_{n,m}\,\frac{[\eta(t)]^n}{[1+\eta(t)]^{n+1}}.
\eea

\end{document}